\documentclass[fleqn,usenatbib]{mnras}

\usepackage[T1]{fontenc}

\DeclareRobustCommand{\VAN}[3]{#2}
\let\VANthebibliography\thebibliography
\def\thebibliography{\DeclareRobustCommand{\VAN}[3]{##3}\VANthebibliography}

\usepackage{graphicx}	
\usepackage{amsmath}	
\usepackage{amssymb}	

\title[Variation mechanism of B2 1633+382]{Interpreting the variation phenomena of B2 1633+382 via the two-component model}

\author[Y.-F. Wang and Y.-G. Jiang]{Yi-Fan Wang,$^{1}$
and Yun-Guo Jiang$^{1}$\thanks{E-mail: jiangyg@sdu.edu.cn}
\\
$^{1}$Shandong Provincial Key Laboratory of Optical Astronomy and Solar-Terrestrial Environment,
Institute of Space Sciences,  \\
Shandong University,  Weihai 264209, China
}

\date{Accepted XXX. Received YYY; in original form ZZZ}

\pubyear{2020}

\begin{document}
\label{firstpage}
\maketitle

\begin{abstract}
Blazars are variable targets in the sky, whose variation mechanism remains an open question. In this work, we make a comprehensive study on the variation phenomena of the spectral index and polarization degree (PD) to deeply understand the variation mechanism of B2 1633+382 (4C 38.41). We use the local cross-correlation function (LCCF) to perform the correlation analysis between multi-wavelength light curves. We find that both $\gamma$-ray and optical $V$-band  are correlated with the radio 15 GHz at the beyond 3$\sigma$ confidence level. Based on the lag analysis,  the emitting regions of $\gamma$-ray and optical locate at $14.2_{-2.4}^{+0}$ pc and $14.2_{-8.3}^{+8.3}$ pc upstream of the core region of radio 15 GHz, and are far away from the broad-line region (BLR). The broad lines in the spectrum indicate the existence of the accretion disk component in the radiation. Thus, we consider the two-component (TC) model, which includes the relative constant background component and the varying jet component to study the variation behaviors. The Markov Chain Monte Carlo (MCMC) procedure is adopted to study the physical parameters of the jet and the background components. To some extent, the study of normalized residuals indicates that the TC model fits better than the linear fitting model. The jet with helical magnetic field is hopeful to explain the variation, and the shock in jet model is not completely ruled out.
\end{abstract}

\begin{keywords}
galaxies: quasars: individual (B2 1633+382) -- galaxies: jets -- $\gamma$-rays: general-- polarization
\end{keywords}



\section{Introduction} \label{sec:intro}

Blazars are {radio-loud} active galactic nuclei (AGNs) with jets aligned close to our line of sights. The remarkable {characteristic} of blazars is their {variabilities} of the flux, spectral index, and polarization.
The variation mechanism of blazars is still under debate. {Many} historical works have {postulated} physical models to explain the variation {phenomena}.  The internal shock in jet {model}, which considers the different velocities of the fluid shells, {was} widely adopted to interpret the {non-thermal} emission of jet \citep{Marscher:1985, Bottcher:2010}.
Some blazars show the quasi periodicity for the radio and optical light curves, which can be interpreted by the helical motion of emitting blob \citep{Ostorero:2004}. The helical jet model was also suggested to explain the variation of polarization \citep{Raiteri:2017}. Besides, the magnetic reconnection and external source model are alternative mechanisms to explain certain abnormal variations \citep{Giannios:2013,Tavani:2015}.
AGNs contain substructures, for instance the disk, which contribute to the radiation at optical bands. Thus, the {coupling of} multiple parameters of emitting components makes it difficult to reveal the variation mechanism.
A comprehensive study of the multiple variation phenomena is necessary to {disentangle the variables and} constrain the variation mechanism.

B2 1633+382, also known as 4C 38.41, is a typical flat spectrum radio quasar (FSRQ) target \citep{Pauliny:1978,Colla:1973,Pilkington:1965}. Its redshift is 1.814 \citep{Paris:2017}. As one of the brightest $\gamma$-ray blazars, it is an aimed target for many monitoring campaigns.
\citet{Raiteri:2012} collected the multiple bands and polarization data of this target, and suggested that the changed viewing angle in the frame of shock in jet model can account for the dependence of the polarization on the optical brightness.  \citet{Hagen:2019} investigated its variability, and  suggested that the redder when brighter trend for this target can be explained by a constant component plus a redder varying component.
With the Very Long Baseline Interferometry (VLBI) observations, it was evident that the position and luminosity of jet features oscillate \citep{Ro:2018}, and there is a helical jet structure \citep{Algaba:2019}.
Comparing the VLBI with flares, it was indicated that the peaks of two prominent $\gamma$-ray flares coincide with the ejection of two new features at the radio band \citep{Algaba:2018b}. These findings are rich, but it also challenges us  if different variation phenomena could be understood in a consistent and unified manner.

In this work, we collect the {multiwavelength} data of $\gamma$-ray, optical and radio, as well as the optical polarization {from the public archives}. We perform the correlation analysis between them. The significant correlations between light curves at three energy bands indicate their positional sequence in jet. The variation patterns of both the optical and $\gamma$-ray spectral indices are similar and {of non-linear type}. The  broad-lines in the spectrum indicate the presence of accretion disk component, which suggests us to consider the two-component (TC) model. The dependence of optical polarization on the {brightness} shows similar non-linear behavior, which could also be interpreted by the TC model.
It is remarkable that {these three non-linear variation behaviors could be possibly explained by the} modulation of the viewing angle.
The parameters of the jet and background components are obtained by {the combined analysis of the analytical formula and the Markov Chain Monte Carlo (MCMC) procedure}.
These investigations present us with a quantitative method to disentangle parameters of the jet from that of the background, and present a benchmark study on the variation mechanism of FSRQ targets.

\section{Data collection} \label{Sec:data}

The multiwavelength data of the target include the $\gamma$-ray data of the {\it Fermi-LAT} \footnote{\url{https://fermi.gsfc.nasa.gov/ssc}}, the optical data of {both the} Steward Observatory (SO)\footnote{\url{http://james.as.arizona.edu/~psmith/Fermi}.} and the Katzman Automatic Imaging Telescope (KAIT)\footnote{\url{http://herculesii.astro.berkeley.edu/kait/agn}.}, {and the radio 15 GHz} data of Owen Valley Radio Observatory (OVRO)\footnote{\url{http://www.astro.caltech.edu/ovroblazars}.}\citep{Atwood:2009,Smith:2009,Li:2003,Richard:2011}

The Fermi-LAT data {in the energy interval of $0.1$-$300$ GeV}, starting from 2008 August 4 to 2019 July 8,  are analyzed by the {\it Fermitools} (version 1.0.1) distributed in the Conda system. A standard pipeline for {\it unbinned likelihood} analysis is implemented to extract the $\gamma$-ray flux and spectrum \citep{Abdo:2009}. In this pipeline, the {\it Fermi}-LAT 8-year catalog (4FGL, \citealt{Abdollahi:2020}), together with the diffuse Galactic and isotropic backgrounds, i.e., {\it gll\_iem\_v07.fits} and {\it iso\_P8R3\_SOURCE\_V2\_v1.txt}, as well as the instrument response functions {\it P8R3\_SOURCE\_V2}, are considered. The time bin is one week. Two types of energy bins are considered. The first type takes the total energy interval as one bin, aiming to obtain the {energy integrated} $\gamma$-ray light curve. The second type divides the energy range of $0.1$-$219$ GeV into seven logarithmically equal bins, aiming to obtain the spectral index of $\gamma$-ray  by linearly fitting at least four fluxes in one time bin. This produces {relatively high quality data of spectral indices} independent of the reduction procedure of $\gamma$-ray fluxes.  For all cases, we use the condition of test statistics (TS$\ge 10$) to filter the fluxes.

The optical photometry and polarization data are retrieved from the long-term  monitoring program operated by the SO \citep{Smith:2009}. These observations were performed by the 1.54-m Kuiper and the 2.3-m Bok telescopes. In this work, we study the $V$ and $R$ band photometry data, as well as the data of polarization degree (PD) obtained by SO from 2008 October 3 to 2018 July 7. {We also collect the optical spectroscopy data from SO, and study the emission lines for several epochs.}
 The other optical photometry data were obtained by the KAIT at Lick Observatory \citep{Li:2003,Richmond:1993,Treffers:1995}. The observations were performed without filters, and the measured magnitudes were transformed to $R$-band roughly \citep{Li:2003,Stahl:2019}. The strict transformation procedure considers the instrument magnitudes and the color terms of both the standard star and the target \citep{Li:2003}. However, the color term of the target is not considered in the pipeline of the transformation.  The comparison between the KAIT and SO $R$-band data in the same day indicates that magnitude difference between them is less than  0.15 magnitude, which roughly agrees with the predicted calibration error of the color term. The time duration of the KAIT photometry data is from 2011 September 1 to 2019 July 30.

\begin{figure}
 \centering
 \includegraphics[width=\columnwidth]{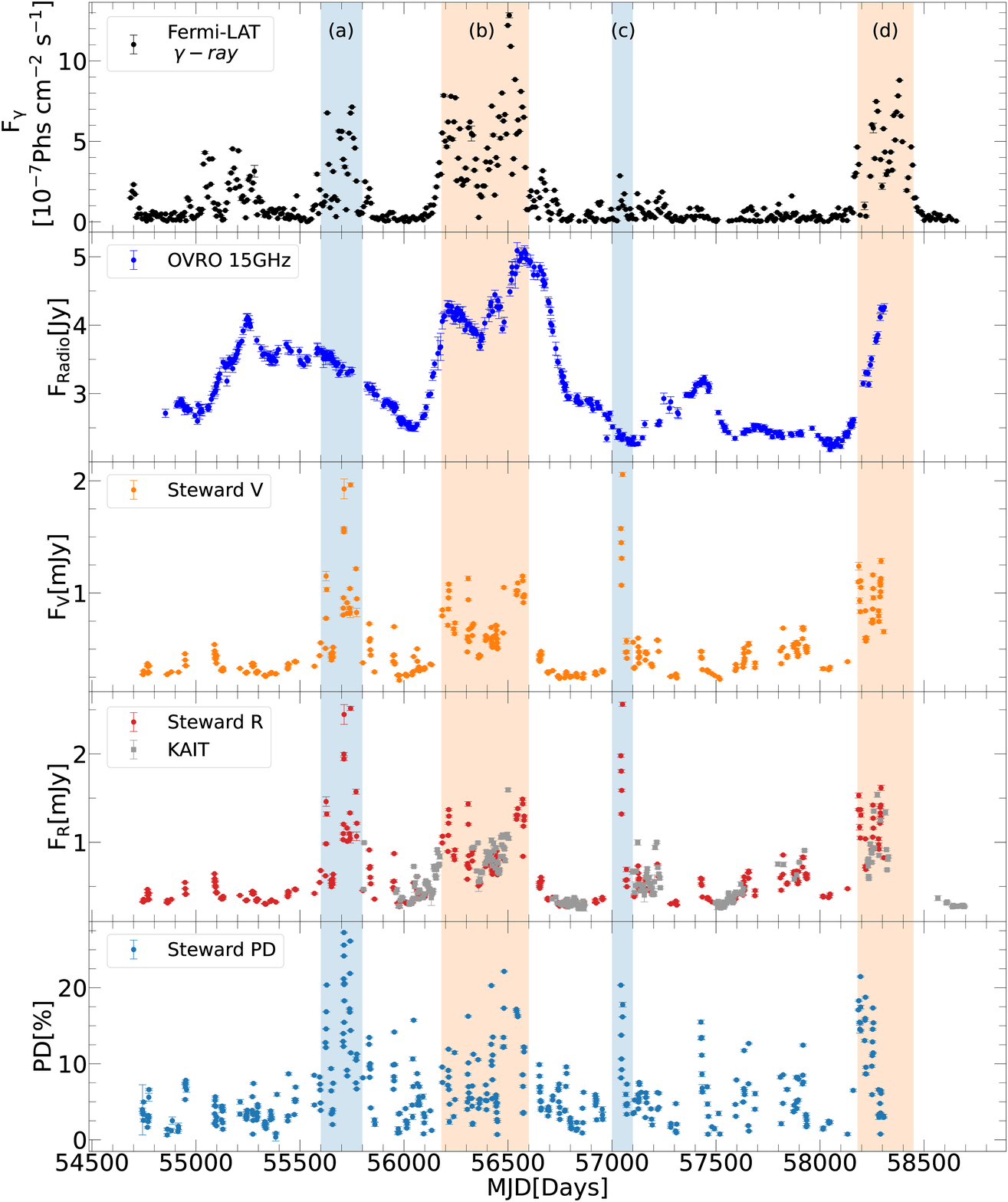}
 \caption{From top to bottom panels, the light curves of $\gamma$-ray ($0.1$-$300$ GeV), radio 15GHz, {optical $V$ band}, optical $R$ band and PD are plotted, respectively. {Four active states of $\gamma$-rays are marked with the colorful shaded areas, and are denoted as period (a), (b), (c), and (d), respectively. The blue and orange color are used to distinguish different states of the radio light curve. }}\label{LC}
\end{figure}

 The data at radio 15 GHz were collected from the OVRO 40-m monitoring program \citep{Richard:2011}. In this work, the calibrated data span from 2009 January 22 to 2020 January 24. In this time duration, 518 data points are present.
The light curves of $\gamma$-ray, radio 15 GHz, {optical $V$ band (Steward)}, optical $R$ band (Steward and KAIT), and optical PD {(Steward)} are plotted in Figure \ref{LC}. {The light curves at all three bands show the flares with periodicity. \citet{Otero:2020} found that there is 1.59 yr periodicity for the $\gamma$-ray light curve. This periodicity may also be true for the optical light curve, since the $\gamma$-ray and optical light curve (Steward data) have many coincident flares by visual inspection. However, the radio peaks are not synchronized with the $\gamma$-ray and optical ones. For optical $R$-band, the monitoring of KAIT misses some flares evident in the Steward light curve. This may lead to some discrepancies in the correlation analysis. The optical PD corresponds to the optical flux well, which indicates a good correlation between them. The more detailed correlation analysis between these variables will be given in the next sections.}

\section{Locations of emitting regions} \label{Sec:lag}
\begin{figure*}
        \includegraphics[width=0.65\columnwidth]{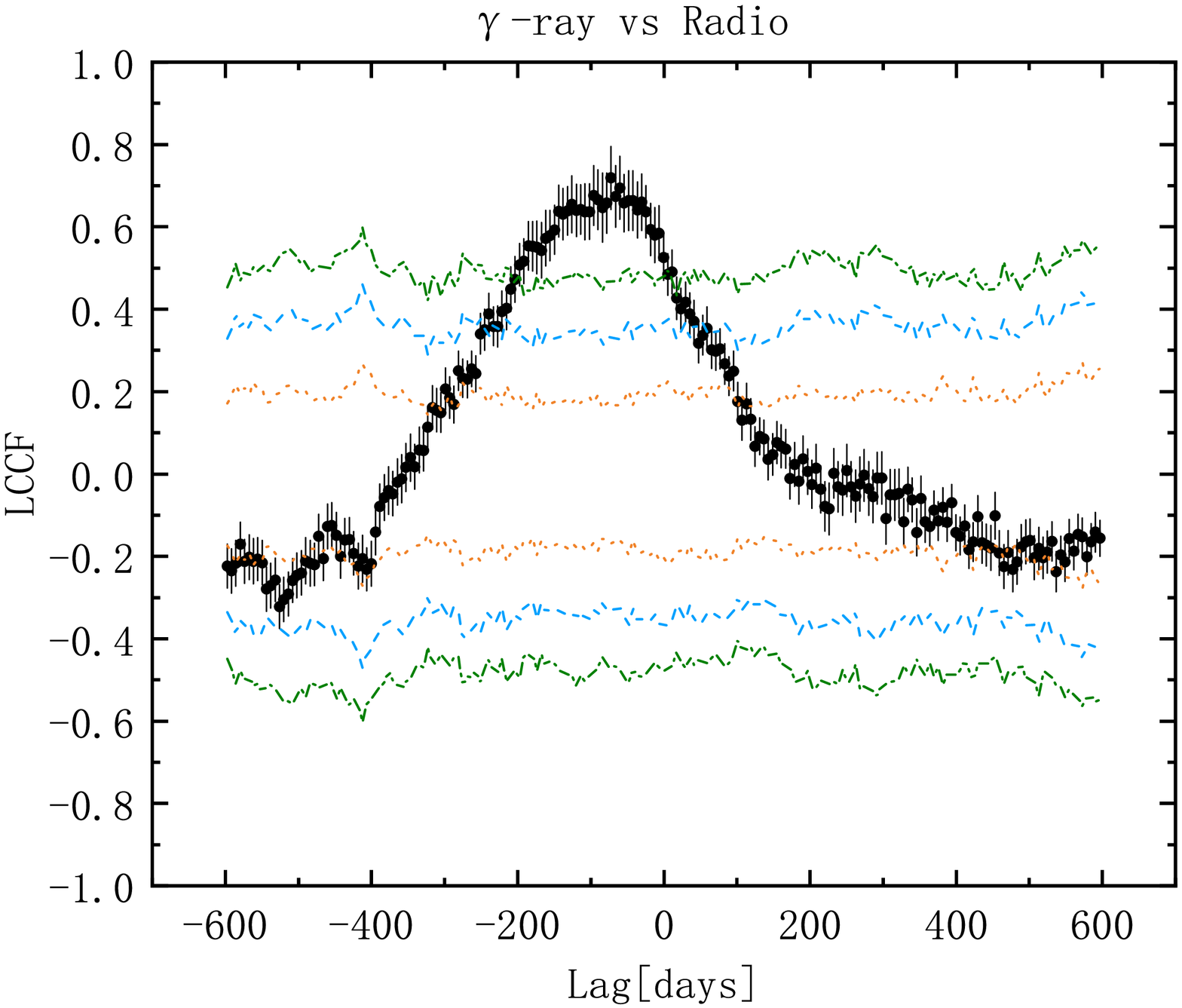}
        \includegraphics[width=0.65\columnwidth]{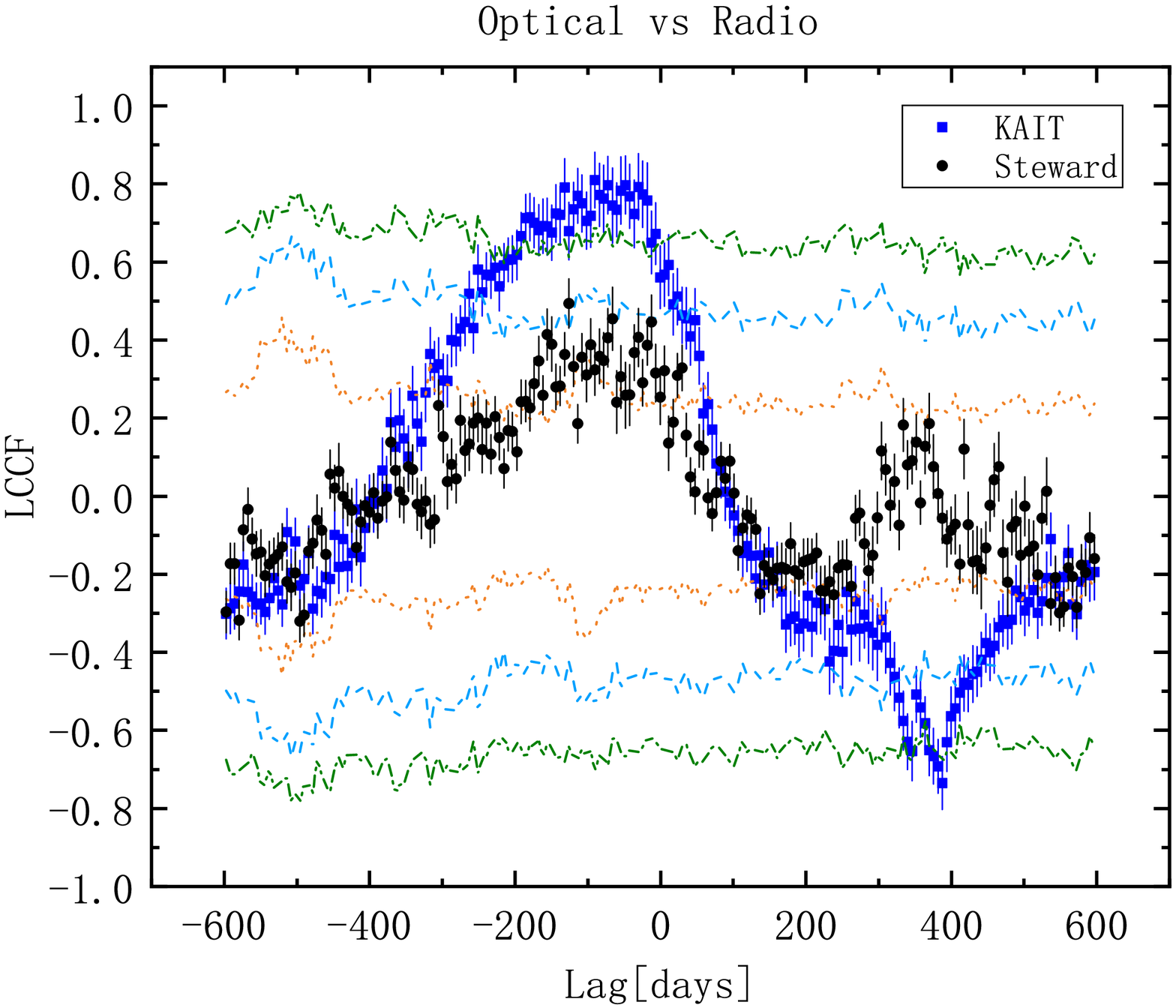}
        \includegraphics[width=0.65\columnwidth]{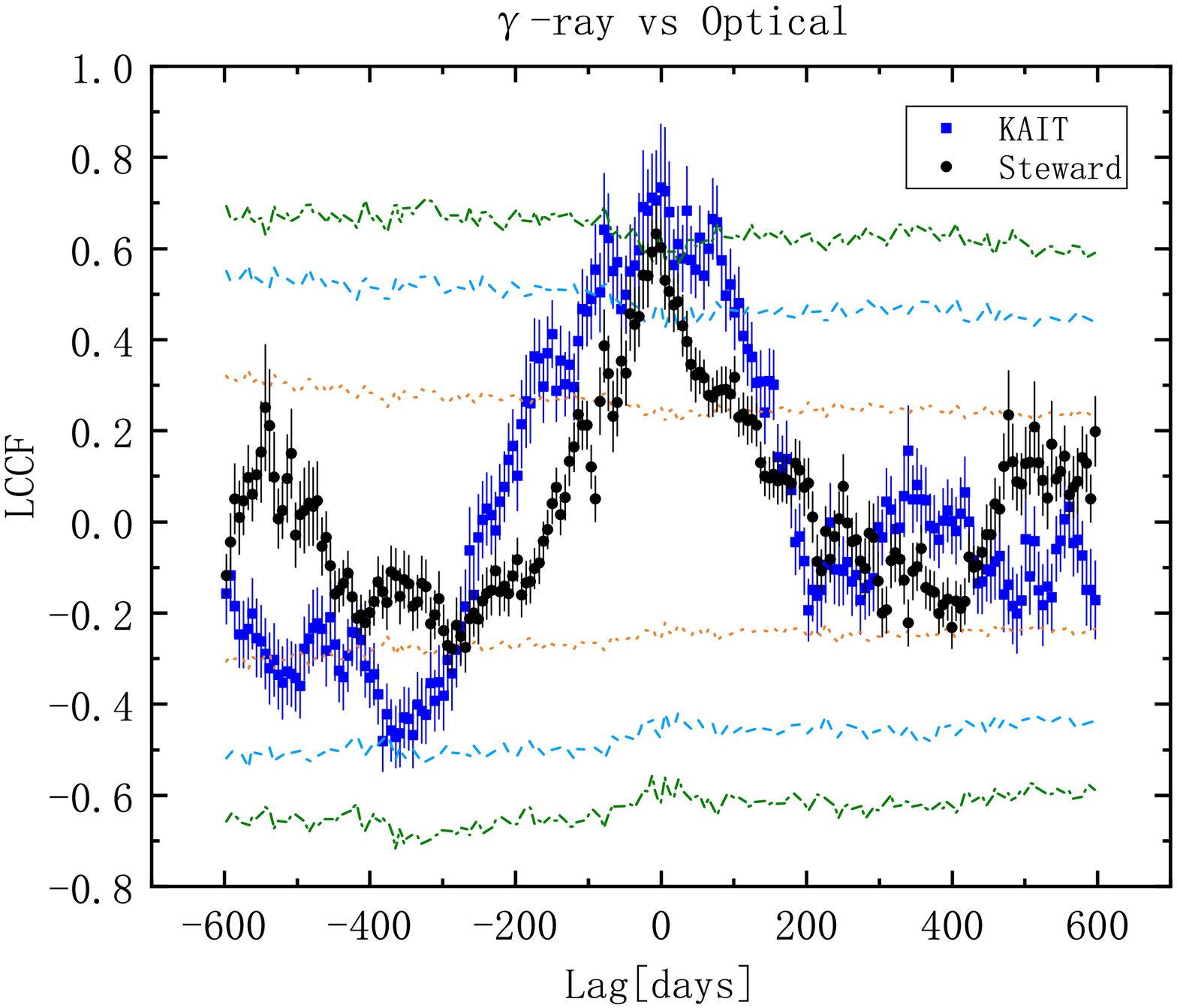}
\caption{The left, middle and right panels present the LCCF results of $\gamma$-ray versus radio (15 GHz), optical ($R$ band) versus radio, and $\gamma$-ray versus optical, respectively. In the LCCF calculation, the $\gamma$-ray in the range of $0.1-300$ GeV is considered. In the center and right panels, the blue squares and black dots denote LCCF results obtained by considering the optical KAIT and Steward $R$ band light curves, respectively. The red dotted, cyan dashed and olive dashed dotted lines denote the $1\sigma$, $2\sigma$ and $3\sigma$ confidence level curves, respectively. {Note that we use the KAIT sampling (rather than the Steward sampling) to obtain the significance lines in the center and right panels.}  The negative lag indicates that the former leads the latter. }\label{Fig:LCCF}
\end{figure*}

Firstly, the cross-correlation analyses between different light curves were performed by using the local cross-correlation function (LCCF, \citealt{Welsh:1999,Max:2014a}). The LCCFs of $\gamma$-ray ($0.1$-$300$ GeV) and optical $R$-band versus radio 15 GHz are plotted in Figure \ref{Fig:LCCF}. The range of lag time is taken to be $[-600, 600]$, {and the lag bin is 6 days.} The confidence level curves are evaluated by the Monte Carlo (MC) simulation \citep{Shao:2019,Jiang:2020}. We simulate $10^4$ artificial light curves by the method described in \citet{Timmer:1995} (TK95). The slopes of the power spectral density (PSD) are set to be $\beta_{\rm radio}=2.1$ and $\beta_{\gamma}=1.5$ \citep{Max:2014a}. There are $10^4$ data points with the bin of one day in each simulated light curve. We also consider the uneven sampling effect in our simulation and extract the subsets of data points with the exact same sampling of the observed radio and TS filtered $\gamma$-ray light curves.  Then the confidence levels of 1$\sigma$, 2$\sigma$, and 3$\sigma$ corresponding to the chance probability of 68.26\%, 95.45\%, and 99.73\% are obtained by calculating LCCFs between the subsets and the observed light curves. In Figure \ref{Fig:LCCF}, the $\gamma$-ray, KAIT optical, and radio 15GHz  light curves are mutually correlated {at beyond the 3$\sigma$ confidence level curve. }
 However, {the LCCF of Steward optical versus radio is less significant than that of KAIT versus radio. This is possibly due to the sampling effect. During periods (a) and (c) (marked with light blue vertical stripes), no radio peaks correspond to optical peaks of Steward sampling, also the KAIT sampling did not monitor these optical peaks. This leads the significant correlation between KAIT and radio and the less significant correlation between Steward and radio.} The correlation of $\gamma$-ray versus Steward { seems to be less significant than that of $\gamma$-ray versus KAIT. This is probably due to that  orphan flares monitored by the Steward in periods (a) and (c) are absent in the KAIT sampling}. Such mutual orphan flares are common in multi-wavelength light curves of blazars \cite{Liodakis:2018}. The  optical flares monitored by Steward have typical duration time of several days, which is less than those time steps of both the $\gamma$-ray and  radio data. This is another plausible reason. Other possibilities like the multiple emitting regions with small size in jet model can not be excluded by the current data. {In the middle panel of Figure \ref{Fig:LCCF}, we note that an anti-correlation signal of $3 \sigma$ significance at the lag of about 400 days is presented in the KAIT versus radio case. This significant anti-correlation is most probably caused by the characteristic (the quasi-periodicity) of the KAIT light curve and has no physical meaning. }

We evaluate the time lag and its 1$\sigma$ standard deviations through MC simulation known as the flux redistribution/random subset selection (FR/RSS) procedure \citep{Peterson:1998, Raiteri:2003}. This procedure was performed $10^4$ times. Two kinds of time lag $\tau_p$ and $\tau_c$ are considered. $\tau_p$ is defined as the lag for the highest peak of LCCF, while $\tau_c$ is the centroid lag defined as $\tau_{c}\equiv \sum_{i} \tau_{i} C_{i} / \sum_{i} C_{i}$, where $C_i$ is the coefficient satisfying $C_i >0.8{\rm LCCF} (\tau_p)$. All the time lags are summarized in Table \ref{timelag}.
One obtains that the $\gamma$-ray leads radio with $\tau_p=-72^{+12}_{-0}$ and $\tau_c=-84^{+6}_{-8}$ days, while the KAIT $R$-band leads radio with $\tau_p=-72^{+42}_{-42}$ and $\tau_c=-97^{+11}_{-10}$ days. {The zero error of the lag is due to the bin effect and the non-Gaussian distribution of lags.} The $\gamma$-ray leads KAIT $R$-band with $\tau_p=-0.5^{+0}_{-12}$ and $\tau_c=4^{+9}_{-10}$ days.
Within uncertainties, we conclude that the optical and  $\gamma$-ray emitting regions are the same, and locate at the upstream of the radio emitting region in jet.
Using the discrete correlation function (DCF), \citet{Algaba:2018a} performed the lag analysis for this target and obtained that $\gamma$-ray and optical lead 15 GHz radio by $67 \pm 40$ and $41 \pm 20$ days, respectively. \citet{Cohen:2014} presented that $\gamma$-ray leads optical by $\tau_c\sim5.9 \pm 0.8$ days using DCF analysis, whose peak significance value is 87.1\%.
\citet{Zhang:2017} presented that the optical leads radio by $\tau_c\sim35$ days with peak DCF$\sim0.68$. These arguments roughly agree with our results in Table \ref{timelag}. The time series with higher cadence are needed to improve the uncertainty of lag.

\begin{table*}
\begin{minipage}{88mm}

\caption{Time lags and relative distances.}
\label{timelag}
\begin{tabular}{lcccc}
\hline
\hline
Pairs &\multicolumn{2}{c}{Time lags (days)} & \multicolumn{2}{c}{Distance (pc)}\\
\cline{2-3} \cline{4\,-5}
 & $\tau_p$ & $\tau_c$ & $ D_p$ & $ D_c$\\
\hline
$\gamma$-ray vs radio			& $-72^{+12}_{-0}$ & $-84^{+6}_{-8}$ & $14.2_{-2.4}^{+0}$ & $16.6_{-1.2}^{+1.6}$ \\
Optical vs radio		        & $-72^{+42}_{-42}$ & $-97^{+11}_{-10}$ & $14.2_{-8.3}^{+8.3}$ & $19.2_{-2.0}^{+2.0}$ \\
$\gamma$-ray vs optical			& $-0.5^{+0}_{-12}$ & $4^{+9}_{-10}$ & $0.1_{-0.0}^{+2.4}$ & $-0.8_{-1.8}^{+2.0}$ \\
\hline
\end{tabular} \\
{Here $\tau_p$ and $\tau_c$ denote the peak and centroid time lags (in unit of days), respectively. The negative lags indicate that the former leads the latter. $D_p$ and $D_c$ are distances derived according to $\tau_p$ and $\tau_c$, respectively.}
\end{minipage}

\end{table*}

The time lags between different bands can be well explained by the jet model. The disturbance propagates along the jet, and its upstream and downstream emit $\gamma$-ray and radio radiation, respectively. Then the distances between the emitting regions relative to that of a given frequency are expressed as \citep{Kudryavtseva:2011,Max:2014a}
\begin{equation}
\Delta \mathrm{D}=\frac{\beta_{\rm app} c \Delta T}{(1+z) \sin \theta},
\end{equation}
where $\beta_{\rm app}$ is the apparent velocity, $c$ is the speed of light, $\Delta T$ is the time lag ($\tau_p$ or $\tau_c$), $z$ is the redshift, and $\theta$ is the viewing angle between the jet axis and the line of sight.
For this target, $\beta_{\rm app}=30$ is adopted from the fastest knot feature \citep{Lister:2019}, and the viewing angle is $\theta\sim2.6^{\circ}$ \citep{Hovatta:2009}. The derived relative distances, $D_p$ and $D_c$, are summarized in Table \ref{timelag}. Based on $\tau_p$, the $\gamma$-ray and optical emitting regions locate at $14.2_{-2.4}^{+0}$ parsec (pc) and $14.2_{-8.3}^{+8.3}$ upstream of the 15 GHz radio core region, respectively.  \citet{Pushkarev:2012} presented the distance from the jet base to the $15$ GHz radio core $r_{\rm core, 15.4 GHz}= 40.67$ pc for this target by using the core shift measurement. If one takes this value, the $\gamma$-ray and the optical emitting regions are about 26.5 pc away from the jet base, that are beyond the BLR.

\section{Variation analysis} \label{Sec:variable}
\subsection{Lines and disk component}
\begin{figure}
\begin{center}
  \includegraphics[width=\columnwidth]{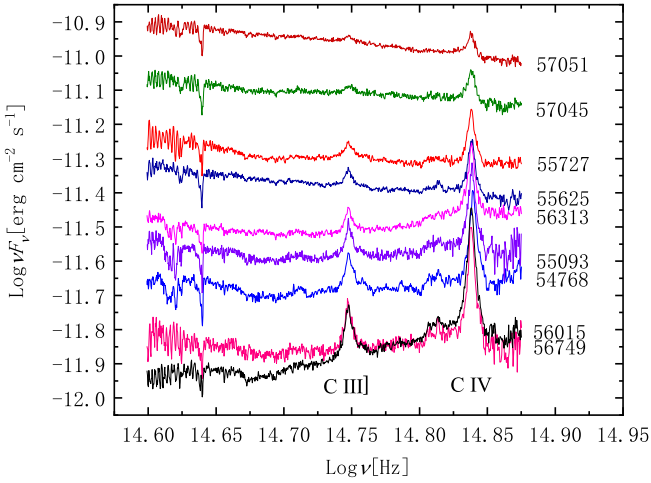}
\caption{ Illustration of the spectra at different epochs for B2 1633+382.  The numbers in the right denote the MJD days.}\label{Spectrum}
\end{center}
\end{figure}

The optical spectra of the target were plotted in Figure \ref{Spectrum}. In that, two prominent broad emission lines can be identified, especially in the low flux state.  By using the {\it SPLAT-VO} \footnote{\url{http://star-www.dur.ac.uk/~pdraper/splat/splat-vo}}, we subtract the continuum and perform the Gaussian fitting of C IV $\lambda$1550 and C III] $\lambda$1909 profiles.  The full width at half maximum (FWHM) of C IV and C III] are obtained as $4620\pm130\ \rm km \ s^{-1}$ and $4760\pm170\ \rm km\ s^{-1}$, respectively. According to the Mg II feature obtained from the SDSS spectrum, we estimate the mass of supermassive black hole (SMBH) and the luminosity of the accretion disk as $M_{\bullet}=10^{9.5}M_{\odot}$ and $L_{\rm acc}=10^{46.6}\rm erg\ s^{-1}$, respectively \citep{Zamaninasab:2014}. {In the standard model of AGN, the broad-lines are produced by the photoionization process of clouds deep in the gravitational potential. The continuum from the accretion disk is widely adopted to be the source of ionization. It was also suggested that the non-thermal emission from the jet can lead to the ionization of BLR. \citet{Arshakian:2010}  proposed that the continuum from the jet can ionize the outflow material around the jet, which was named as the non-virial BLR. For the FSRQ target 3C 454.3, \citet{Leon:2013} found a link between the broad-line fluctuation with the feature traversing through the radio core, which indicates the presence of BLR surrounding the radio core. These unusual BLRs indicate the complexity of the background component. In this work, we assume that the BLR is illuminated by the disk component, and other possibilities are not considered.}

Thus, the observed fluxes contain components of the accretion disk, BLR, and jet. It is obvious that the spectral slope of the high flux state is negative, while that of the low flux state is positive. The synchrotron peak frequency of the source is given as  $\log\nu_p=12.76$ \citep{Zhang:2017}, which is at the near-infrared band. Therefore, if the jet component dominates the observed flux, the spectral slope should be negative (see Figure 6 in \citet{Zheng:2017}). The spectral slope of the low flux state is positive, indicating that the accretion disk component dominates over the jet component at this phase. \cite{Raiteri:2012} found that emission lines are not highly polarized, which indicates the existence of the unpolarized background disk component. Therefore, the TC model should be considered in the investigation of the variation mechanism.

\subsection{The two-component model}
 The observed flux in the TC model contains two components, i.e.,
\begin{equation}  \label{eq:Flux}
F_{\rm total}=F_{\rm jet}+F_{0},
\end{equation}
where $F_{\rm jet}=\delta^{3} F_{\nu^{\prime}}^{\prime}(\nu^{\prime})$ is the jet component, and $F_0$ is a constant component representing the invariant background.  $\delta \equiv[\Gamma(1-\beta \cos \xi)]^{-1}$ is the Doppler factor, and $\xi$ is the viewing angle. {The observed fluxes at the optical $R$ and $V$ bands are expressed as $F_R= F_{{\rm jet,}R}+F_{0,R}$ and $F_V=F_{{\rm jet},V}+F_{0,V}$.}  By tuning $\delta$, the dependence of $\alpha_{\rm opt}$ on $\log F_{V}$ can be studied numerically.
Considering two energy bands $E_1$ and $E_2$ ($E_2>E_1$), the $\gamma$-ray spectral index can be expressed as $\alpha_{\gamma}=\log [F_{\gamma}(E_{2})/F_{\gamma}(E_{1})]/ \log [E_{2}/E_{1}]$.
The fluxes at $E_1$ and $E_2$ are given as $F_{\gamma}(E_{1})=\delta^{3} a+a_{0}, F_{\gamma}(E_{2})=\delta^{3} b+b_{0}$, where $a$ and $b$ denote fluxes emitted from the jet in the comoving frame, and $a_0$ and $b_0$ denote the constant background component. 
By using the Stokes parameters $I$, $Q$, and $U$, PD is given as $\Pi=\sqrt{Q^{2}+U^{2}} / I$. Setting the Stokes parameters of jet component as $I_1$, $Q_1$, and $U_1$, that of the accretion disk component as $I_2$, $Q_2$, and $U_2$, the synthesized PD is written as $\Pi_{\text {total}}=\sqrt{Q_{1}^{2}+U_{1}^{2}}/(I_{1}+I_{2})$,
where we have considered $Q_2=U_2=0$.
{\cite{Raiteri:2013} have considered $\Pi_{\rm jet}$ for two models, whose empirical relations are $\Pi_{\rm jet,H}=P_{\rm max}\sin^2 \xi^{\prime}$ for the jet with helical magnetic field model (denoted as TCH model hereafter) and $\Pi_{\rm jet,S}=P_{0} (1-\eta^{-2})\sin^2 \xi'/[2-(1-\eta^{-2}\sin^2 \xi')]$ for the transverse shock in jet model (denoted as TCS model hereafter), respectively. Here $\xi'$ is the viewing angle in the comoving frame, and  $P_{\rm max}$ and $P_0$ can be considered as the intrinsic PD for synchrotron radiation in jet. The relation of $\Pi_{\rm jet, H}=P_{\rm max}\sin^2 \xi^{\prime}$ was proposed according to cases of diffuse and reverse-diffuse pinch magnetic fields in \citet{Lyutikov:2005}. This relation is generalized to $\Pi_{\rm jet,H}=P_{\rm max}\sin^n \xi^{\prime}$ in \citet{Shao:2019}. By investigating the case of unresolved hollow jet with cylindrical shells, we find that the empirical relation with $n=3$ can well approximate the numerical result (unpublished work).  Thus, we will consider the relation $\Pi_{\rm jet,H}=P_{\rm max}\sin^3 \xi^{\prime}$.
The synthesized (or observed) total PD  can be rewritten as}
\begin{equation} \label{eq:PD}
\Pi_{\rm total}=\Pi_{\rm jet}\frac{F_{\rm jet}}{F_{\rm total}},
\end{equation}
where $\Pi_{\rm jet}$ takes either $\Pi_{\rm jet,H}$ or $\Pi_{\rm jet,S}$. Equations (\ref{eq:Flux}) and (\ref{eq:PD})
will be used to study the variation behavior in  the plot of PD versus fluxes.

\subsection{Variation behaviors and model fittings}
\begin{figure*}
       \begin{minipage}[t]{0.32\linewidth}
        \centerline{\includegraphics[width=\columnwidth]{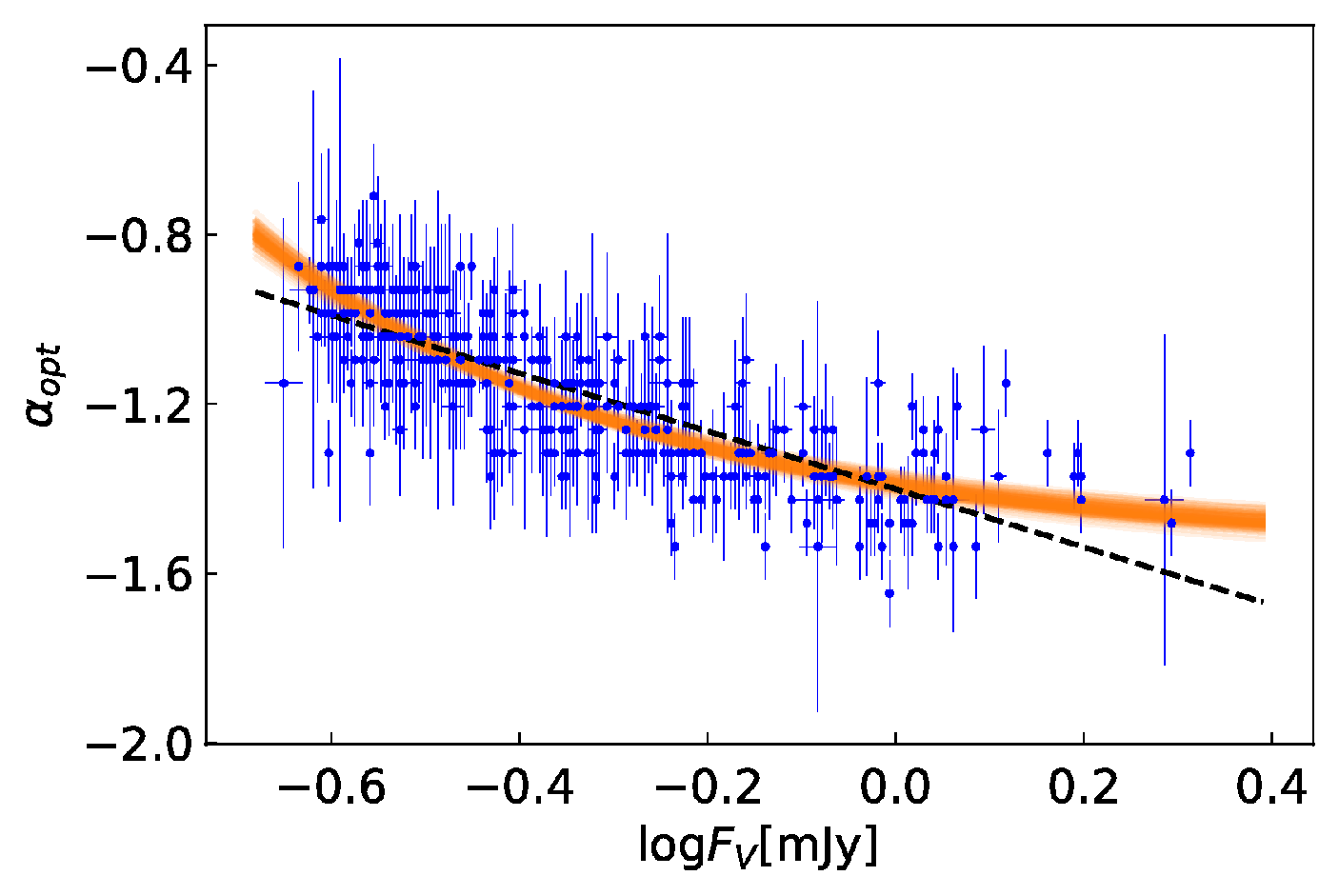}}
        \centerline{(1)}
           \end{minipage}
    \begin{minipage}[t]{0.32\linewidth}
        \centerline{\includegraphics[width=\columnwidth]{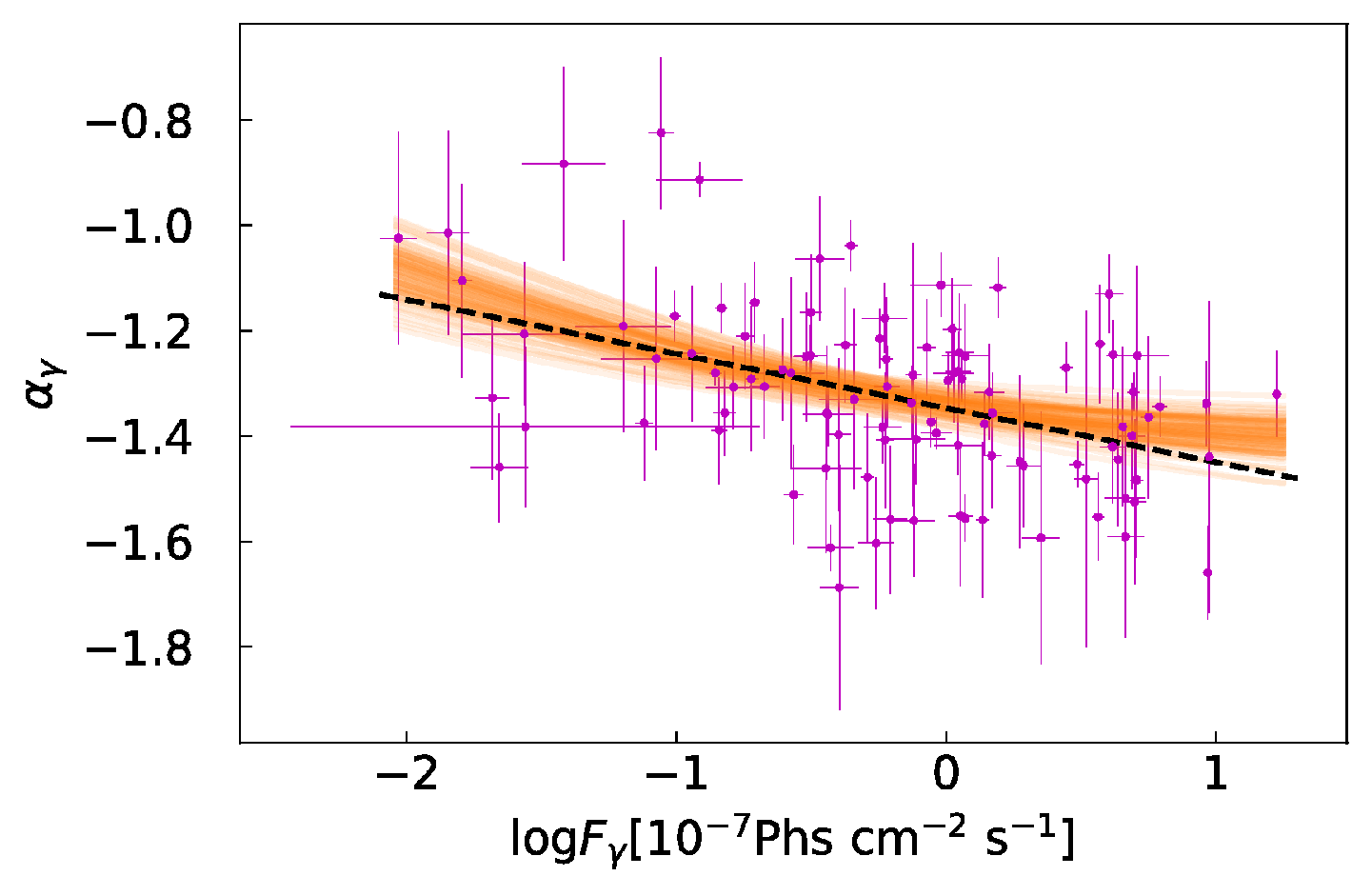}}
        \centerline{(2)}
           \end{minipage}
    \begin{minipage}[t]{0.32\linewidth}
        \centerline{\includegraphics[width=\columnwidth]{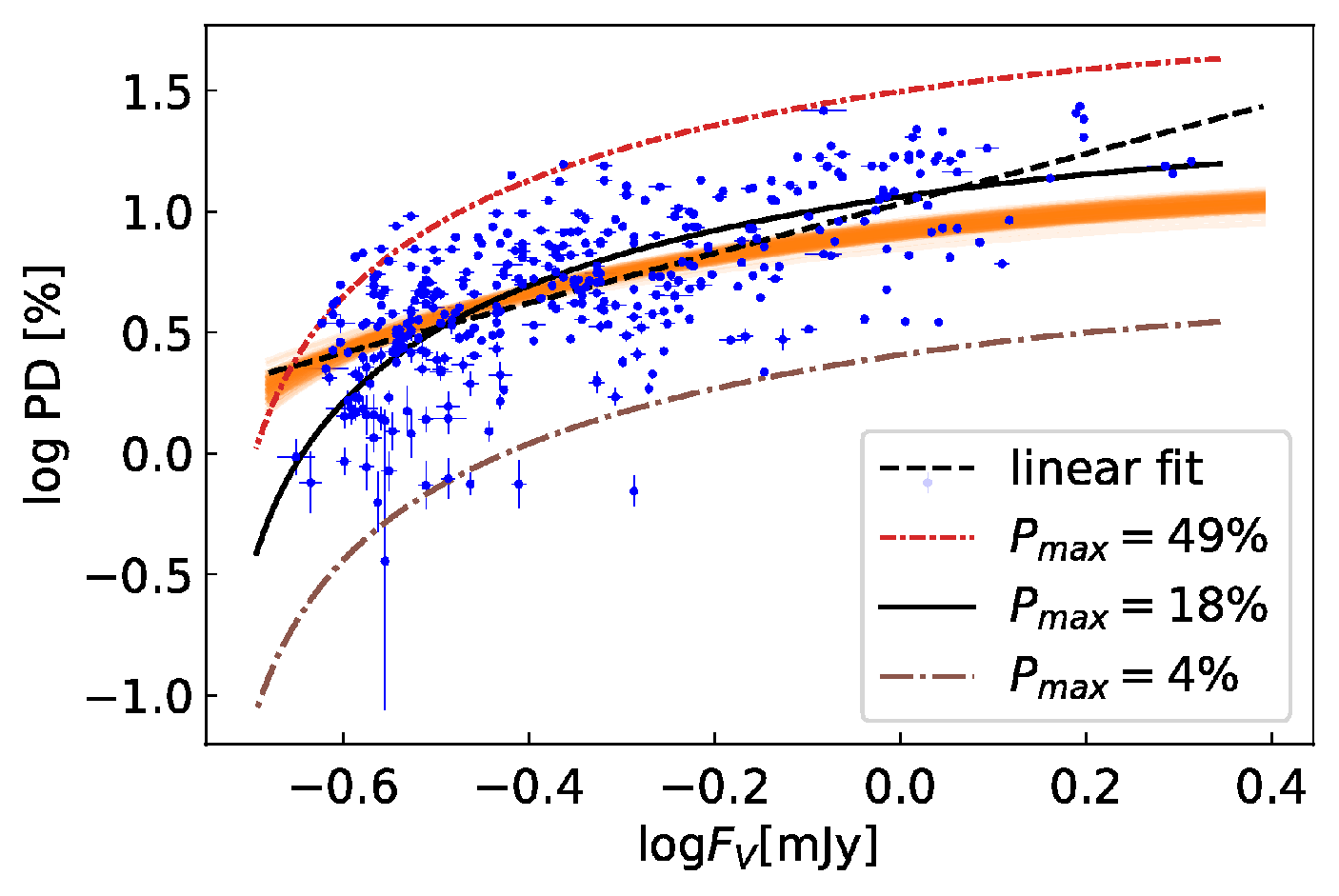}}
        \centerline{(3)}
           \end{minipage}

       \begin{minipage}[t]{0.32\linewidth}
        \centerline{\includegraphics[width=\columnwidth]{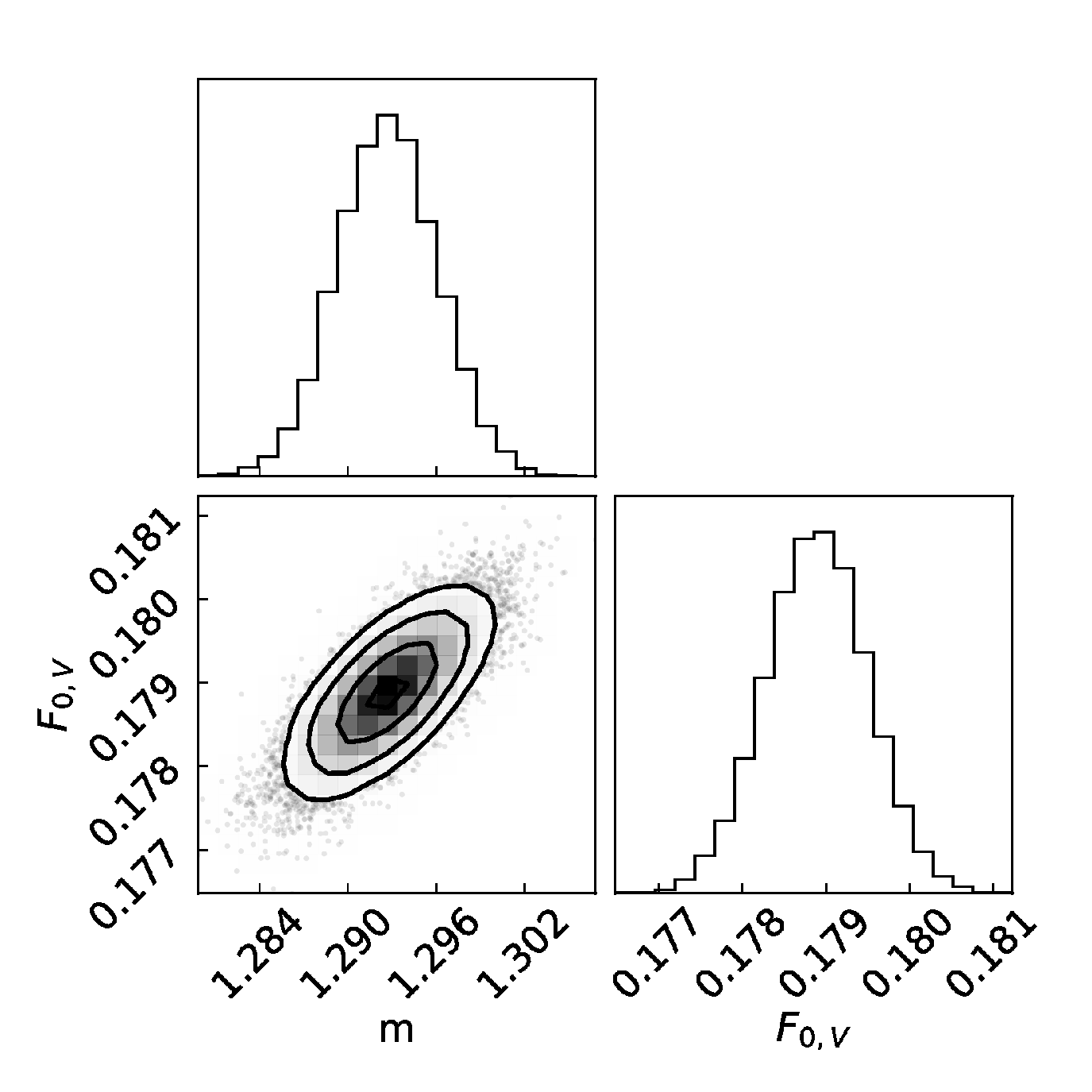}}
        \centerline{(a)}
           \end{minipage}
    \begin{minipage}[t]{0.32\linewidth}
        \centerline{\includegraphics[width=\columnwidth]{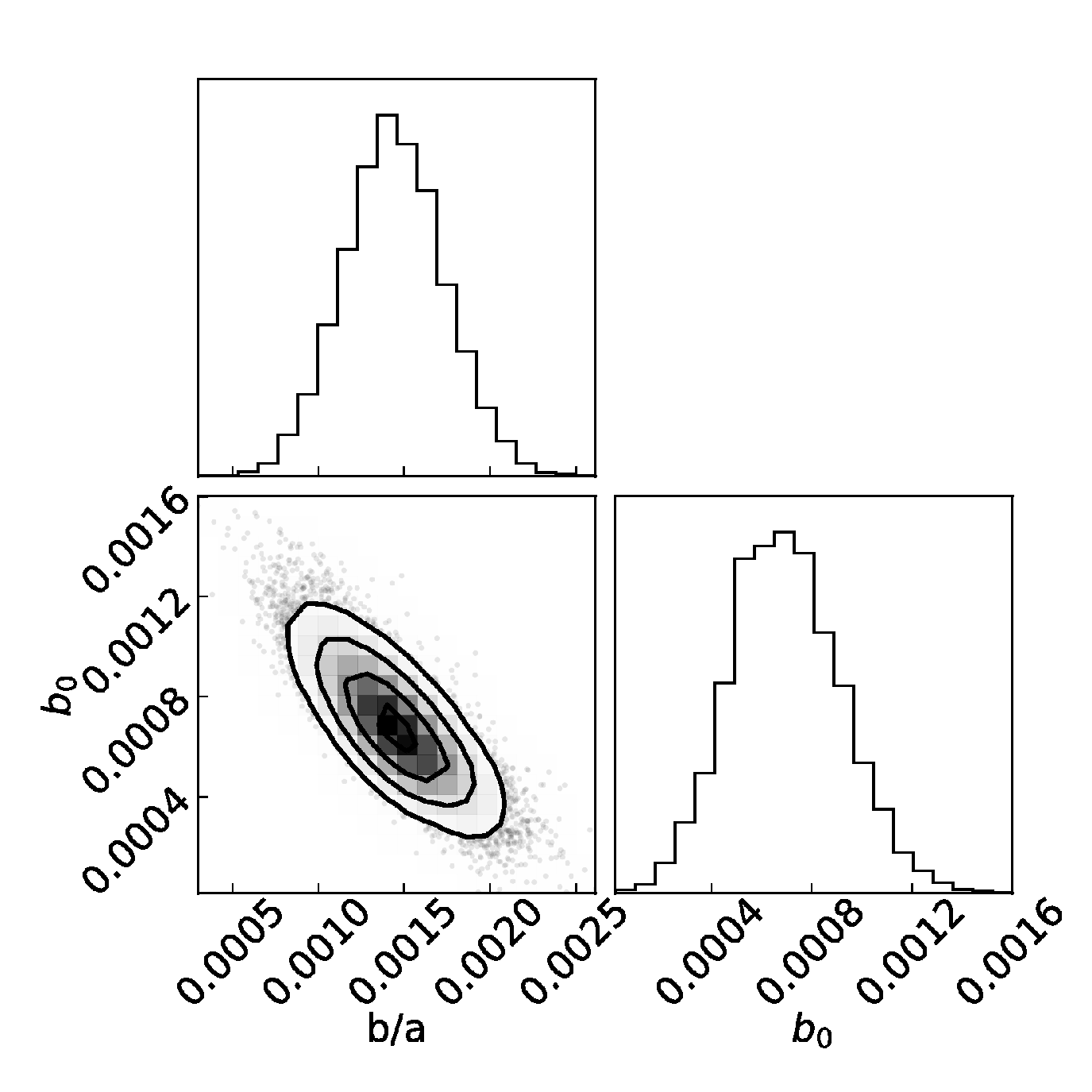}}
        \centerline{(b)}
           \end{minipage}
    \begin{minipage}[t]{0.32\linewidth}
        \centerline{\includegraphics[width=\columnwidth]{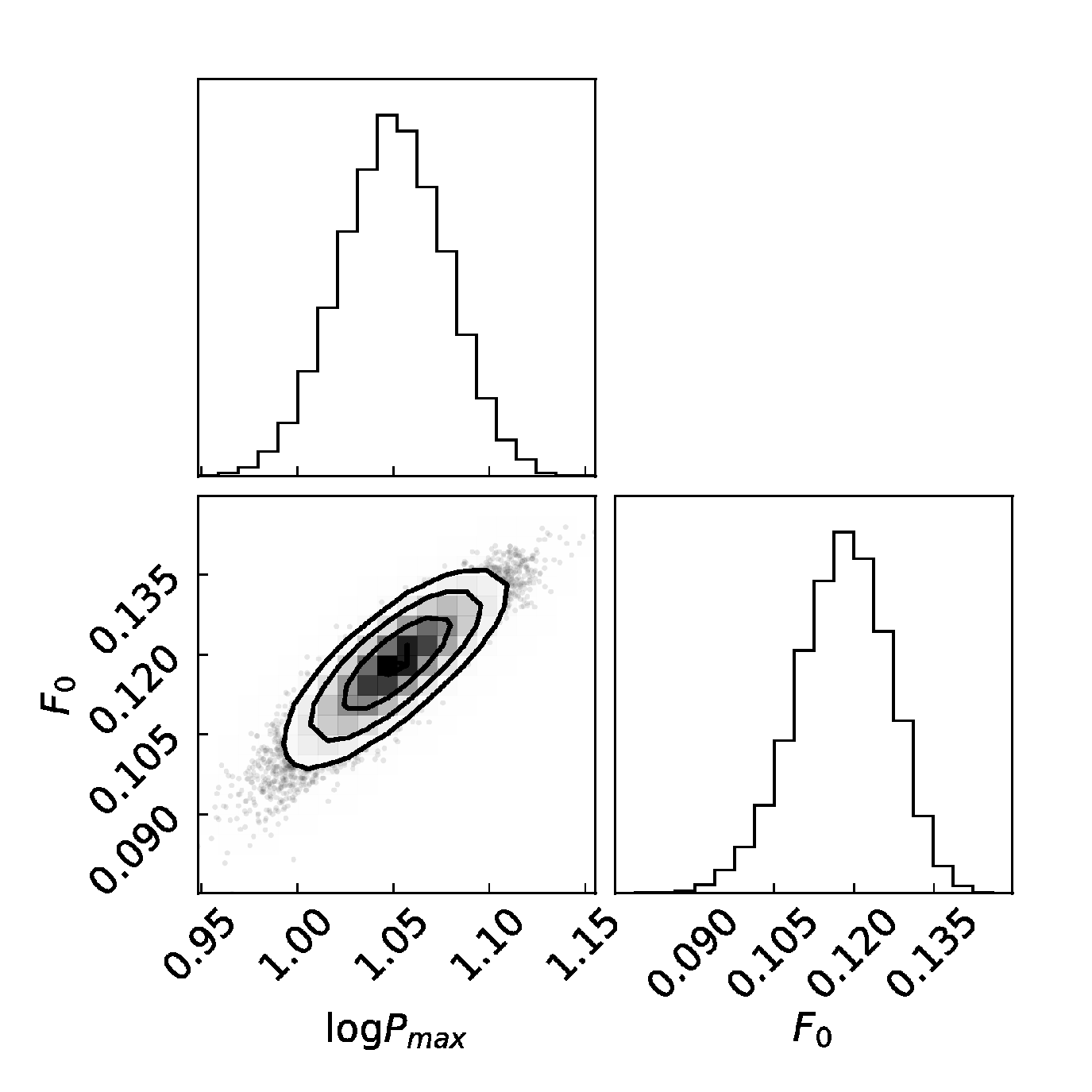}}
        \centerline{(c)}
           \end{minipage}

      \begin{minipage}[t]{0.32\linewidth}
        \centerline{\includegraphics[width=\columnwidth]{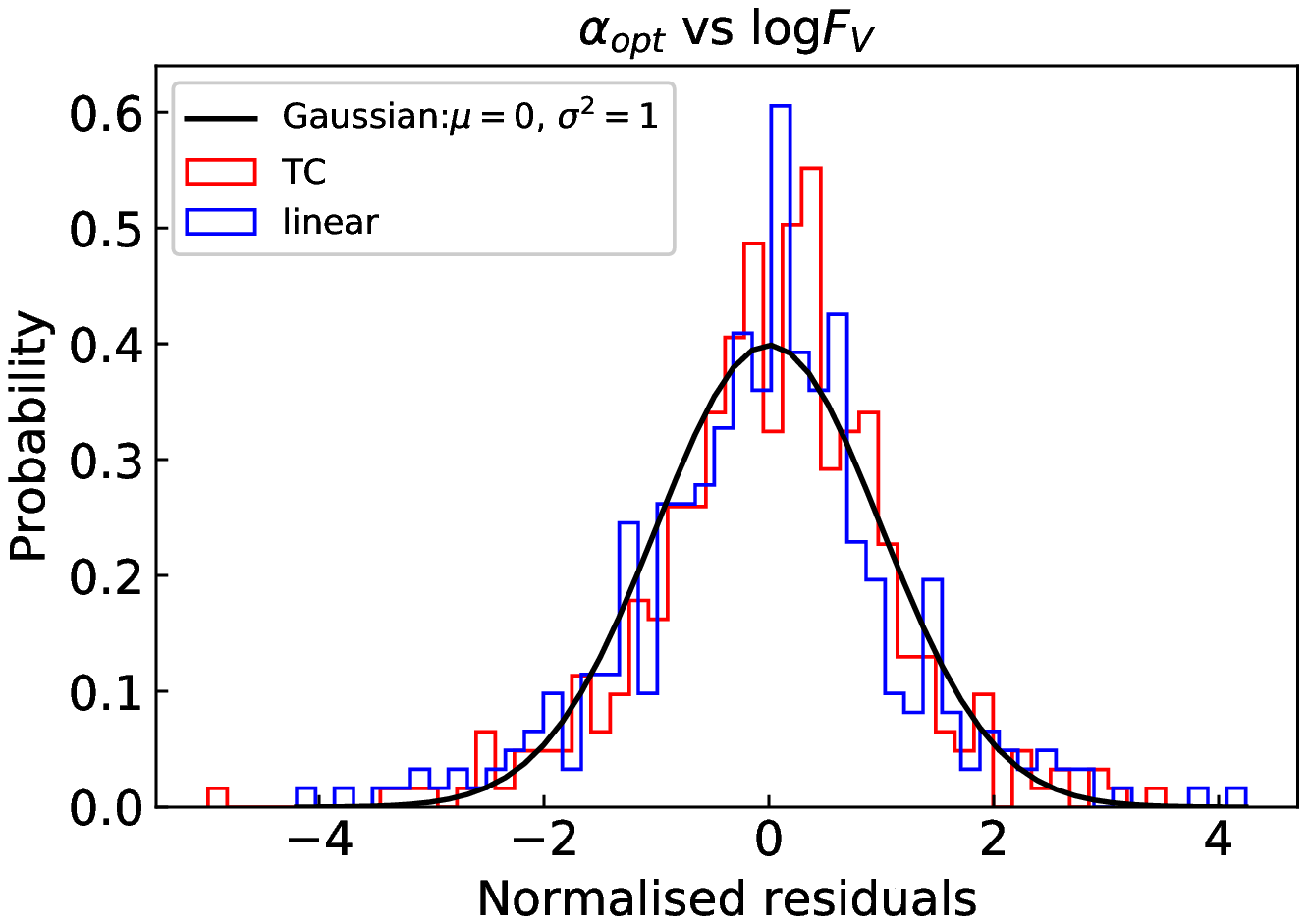}}
        \centerline{(I)}
           \end{minipage}
    \begin{minipage}[t]{0.32\linewidth}
        \centerline{\includegraphics[width=\columnwidth]{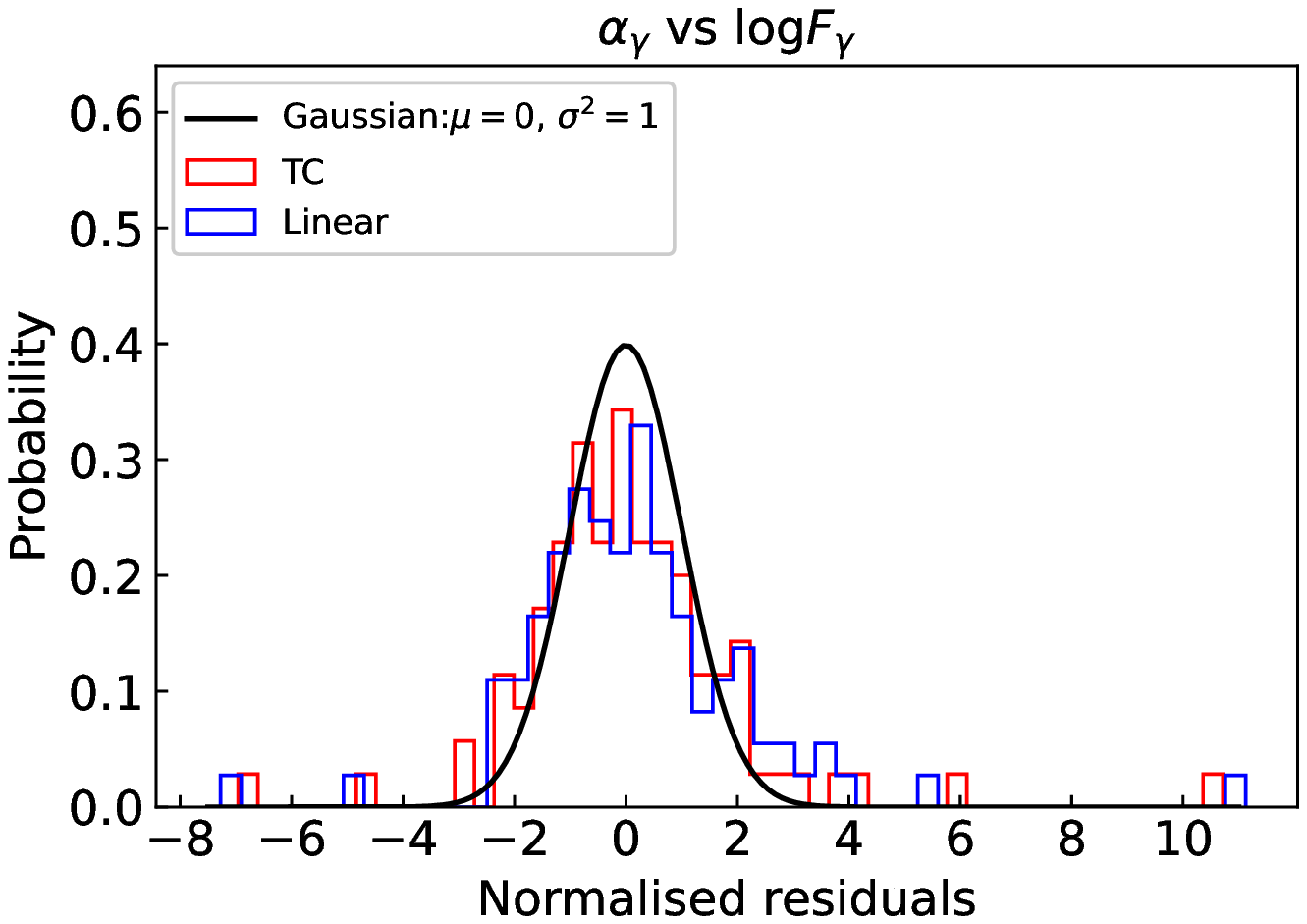}}
        \centerline{(II)}
           \end{minipage}
    \begin{minipage}[t]{0.32\linewidth}
        \centerline{\includegraphics[width=\columnwidth]{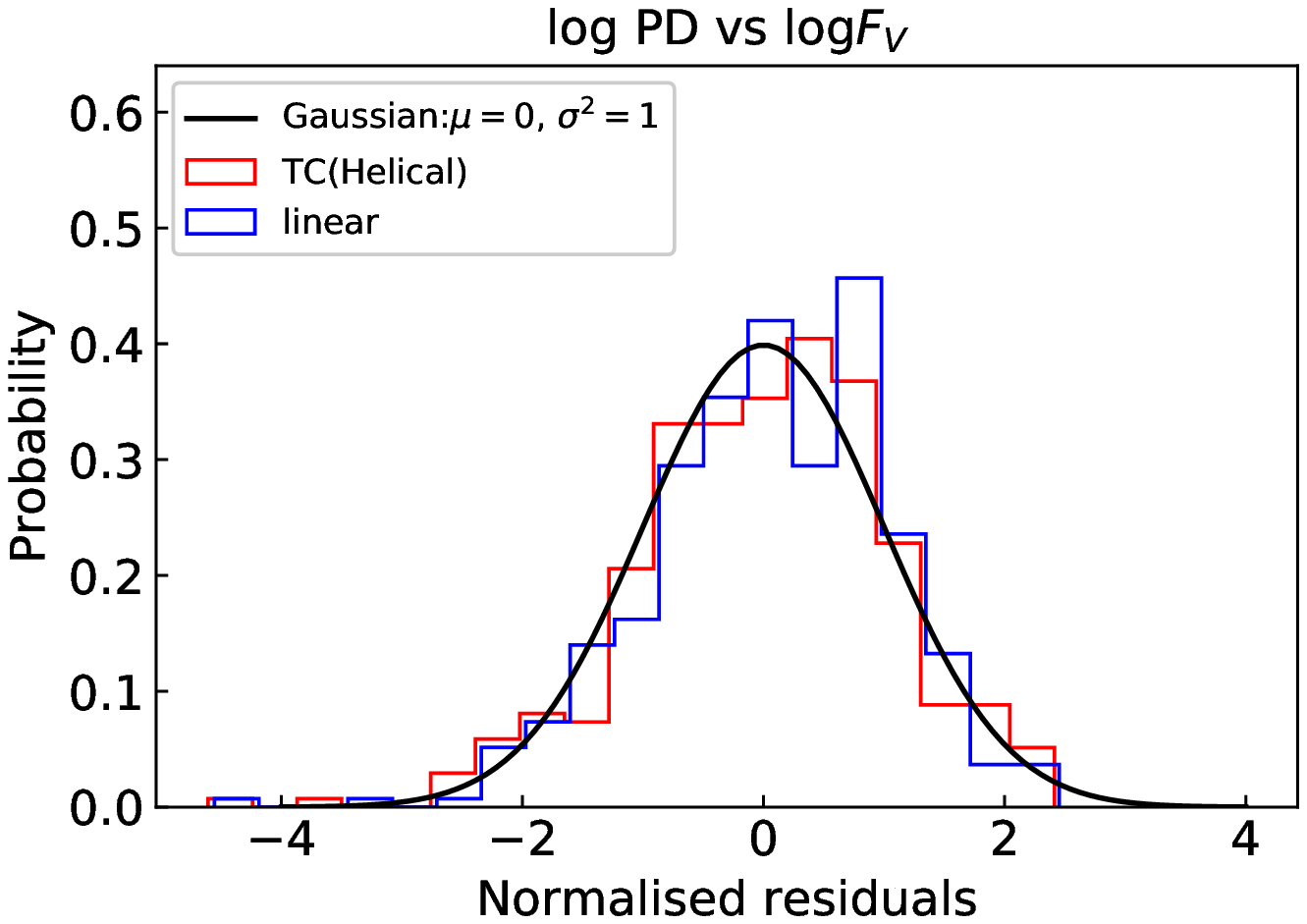}}
        \centerline{(III)}
           \end{minipage}
         \caption{The optical spectral index {$\alpha_{\rm opt}$ versus $\log F_{\nu}$, $\alpha_{\gamma}$ versus $\log F_{\gamma}$ ($0.3$-$0.9$ GeV), and $\log$PD versus $\log F_{V}$} are plotted in the top panels (1), (2), and (3), respectively. The corresponding corner plots of parameters of  TC model are plotted in the {middle} panels (a), (b), and (c), respectively. The histograms in each corner plot indicate the 1-D probability distributions for one parameter obtained by marginalizing over the other parameter. The left bottom plots in each corner plot show the 2-D projections of the posterior probability distributions of each pair of parameters, where the contours indicate the $1\sigma$, $2\sigma$, and $3\sigma$ regions. The orange lines in top panels denote 100 fitting results sampled from the MCMC {procedure, and the black dash lines denote the linear fittings. The distribution of the normalized residuals for all cases are plotted in bottom panels (I), (II), and (III), respectively. The standard Gaussian are represented by the black solid line, and histograms with red and blue colors denote the distributions of NR for the TCH model and linear fitting model, respectively. In panel (III), the red short chain line, the black solid line, and the brown long chain line denote the predictions of TCH model with $P_{\rm max}=49\%, 18\%,$ and $4\%$, respectively.    }
 } \label{Fig:fit}
\end{figure*}

In panel (1) of Figure \ref{Fig:fit}, the spectral index $\alpha_{\rm opt}$ versus $\log F_{V}$  is plotted. {Here, the spectral index is defined as $\alpha_{\rm opt}\equiv \log (F_{V}/F_{R}) / \log (\nu_R/\nu_V)$.} It is evident that $\alpha_{\rm opt}$ decreases as $F_{V}$ increases. This is the softer when brighter (SWB) trend, which is equivalent to the redder when brighter (RWB) trend for the color index behavior. In panel (2), $\alpha_{\gamma}$ (the spectral index of $\gamma$-ray ) versus $\log F_{\gamma}$ ($\gamma$-ray flux in the energy bin of $0.3$-$0.9$ GeV) is presented. A similar SWB trend is evident in this plot. In panel (3), the $\log$PD versus {$\log F_{V}$} is plotted. In general, the PD increases with the flux. These variation behaviors are usually studied by the linear fitting, which is performed with the least square method ({\it scipy.optimize package}), see the black dash lines in panels (a), (b), and (c) of Figure \ref{Fig:fit}.

However, the non-linear fitting, which is the prediction of the TC model, is also hopeful to explain the variation behaviors.
We will implement the Markov Chain Monte Carlo (MCMC) procedure {(the python module {\it emcee})} to figure out the range of input parameters.
Since the Doppler factor $\delta$ is set as the variable, there are four input parameters in the TC model. However, two of them are coupled together, which brings the divergence problem in the MCMC process. Thus, to study the variation of $\alpha_{\rm opt}$, we consider three input parameters, i.e., $F_{0,R}$, $F_{0,V}$ and $m=F_{{\rm jet,}R}/F_{{\rm jet,}V}$.  {To study the variation of} $\alpha_{\gamma}$, the three input parameters are $a_0$, $b_0$, and
$b/a$.  {One can} constrain $F_{0,R}$ by the physical condition that the background constant component should be less than the observed fluxes even in the faintest state. The {observed} highest magnitude of $R$ band is $17.65$, corresponding to   $F_{R}=0.268$ mJy. {Thus, we set $F_{0,R} =0.2$ mJy, and perform the MCMC procedure to obtain $F_{0,V}$ and $m$ in the study of $\alpha_{\rm opt}$ versus $\log F_{V}$. The best fitting results indicate $F_{0,V}=0.179^{+0.001}_{-0.001}$ mJy and $m=1.293^{+0.003}_{-0.003}$.} Varying $F_{0,R}$ in a range of $0.15$-$0.3$ mJy, the fitted results vary moderately at the same order. The distribution of parameters is illustrated by the corner plot in panel (a) of Figure \ref{Fig:fit}. Following the similar argument, for the $\gamma$-ray spectral index {variation} behaviors, we set  $a_0$ to be $10^{-9} {\rm phs \ cm^{-2}\ s^{-1}} $ and $E_2/E_1=100$.  The best fitting parameters obtained from the MCMC procedure are given as $b_0=6.9^{+2.3}_{-2.1} \times 10^{-11} \rm phs \ cm^{-2} \  s^{-1}$ and $b/a=1.4^{+0.29}_{-0.28} \times 10^{-3}$ (see panel (b) of Figure \ref{Fig:fit}).

To fit the PD behavior by using the TC model, we set $F_{\nu^{\prime}}^{\prime}(\nu^{\prime}) = 10^{-4}$ mJy and $\Gamma=31$  \citep{Savolainen:2010}. The variable in this case is $\xi$, which also leads to the variation of $\delta$.  {In the TCH model, both  $P_{\rm max}$ and $F_{0,V}$ are left free in the MCMC.  The best fitting results of MCMC  indicate $ P_{\rm max}=11.22^{+0.75}_{-0.70}$\% and $F_{0,V}=0.118^{+0.008}_{-0.009}$ mJy (see panel (c) of Figure \ref{Fig:fit}). The value of $F_{0,V}$ is just half of that obtained in the study of $\alpha_{\rm opt}$. This inconsistency  brings us a caution for the validity of the MCMC procedure in searching the best fittings for the non-linear model with more than two free parameters. When we consider the TCS model, the input free parameters are $\eta$ and $F_{0,V}$. However, the MCMC procedure is divergent when we consider typical values of input parameters. The possibility that MCMC is convergent with fine-tuned initial values is not ruled out.   }

\subsection{Model assessment}
\citet{Andrae:2010} presented a method to assess the goodness of model fittings. The procedure of the method calculates the normalized residual (NR) first, and compares the histogram of NR with the standard Gaussian. Then, the Kolmogorov-Smirnov (KS) test is used to estimate the goodness of model fittings. In the following, we will use this method to indicate which model is better.

First, the NR for a given model is given as \citep{Andrae:2010}
\begin{equation}
 R_i=\frac{y_i-f(x_i)}{\sigma_i},
\end{equation}
where $f(x_i)$ denotes the model prediction, and $y_i$ and $\sigma_i$ denote the observed value and  measurement error, respectively. Then histogram distributions of NRs for both the TC and linear fitting models are plotted in the bottom panels of Figure \ref{Fig:fit}. If the histogram deviates significantly from  the standard Gaussian ($\mu=0$, $\sigma^2=1$), the model should be ruled out.

Secondly, we recalculate the NR by counting the effect of uncertainty of input parameters. In panel (I) and (II) of Figure \ref{Fig:fit}, the TC models consider the best fitting parameters predicted by the MCMC procedure. If we set $P_{\rm max}=11\%$ and $F_{0,V}=0.118$ mJy (the result of MCMC procedure) to give the prediction of TCH model, the histogram of the TCH model seriously deviates from the standard Gaussian. And the $p$-value of KS-test is of order $10^{-76}$. If this is true, the TCH model seems to be ruled out. However, one assumption for the validity of MCMC is that the free parameters are unique in physics. If the input parameter has the uncertainty, the result of MCMC should be questioned. For TC model in the study of $\alpha_{\rm opt}$, the two free parameters are $F_{0,V}$ and $m$. In physics, $F_{0,V}$ denotes the background flux, and $m$ represents the spectral index of jet emission. It is quite possible that both of them have small uncertainties in physics, that can explain why the TC model fit well to the data. For the TCH model, two free parameters are $P_{\rm max}$ and $F_{0,V}$. It is not a natural condition that $P_{\rm max}$ is unique for all flares of the target. If $P_{\rm max}$ has a large uncertainty, the MCMC results of TCH model is not reliable. To estimate the uncertainty of $\log {\rm PD}$ caused by the uncertainty of $P_{\rm max}$, we study the TCH model manually. Give $F_{0,V}=0.179$ mJy (the prediction of case $\alpha_{\rm opt}$), we calculate the chi-square of the TCH model by trying different $P_{\rm max}$. In panel (3) of Figure \ref{Fig:fit}, the black solid line with $P_{\rm max}=18\%$ has the minimized chi-square value, while the red short chain line with $P_{\rm max}=49\%$  and $P_{\rm max}=4\%$  can roughly fit the upper and lower boundaries of $\log {\rm PD}$. Thus, the uncertainty of $\log {\rm PD}$ is probably due to the uncertainty of $P_{\rm max}$. Using the line with $P_{\rm max}=18\%$, the residuals are best fitted by the Gaussian with $\sigma=0.22$. We consider this standard deviation as a system error, and recalculate the NR by $R_i=(y_i-f(x_i))/(\sigma_i+0.22)$. The histogram of NR is plotted in panel (III) of Figure \ref{Fig:fit}. Up to our knowledge, no quantitative method is available to assess the model with  the uncertainty of input parameters.  Our procedure is based on the  argument of physical setup, and its validity needs further study.


Finally, the output of KS-test, which includes the distance and $p$-values, can  indicate the goodness of model fittings quantitatively. Besides that, we also consider the reduced chi-squares ($\chi^2_{\nu}\equiv \sum_{i}^{N} R_i^2/K$, where $K$ is the degree of freedom) to assess models. The results of KS-test and $\chi^2_{\nu}$ for all cases are listed in Table \ref{tab:ks}.
In the case of $\alpha_{\rm opt}$ versus $\log F_V$, the histograms of NRs in both models are well fitted by the standard Gaussian. The $p$-values indicate that the linear fitting performs better than the TC model, while the $\chi^2_{\nu}$ of TC model is less than that of the linear fitting.  In the case of $\alpha_{\gamma}$ versus $\log F_{\gamma}$, one can conclude that the TC model fits better than the linear fitting based on the KS-test. However, $\chi^2_{\nu}$ is much larger than one. This may be due to that the spectral index of the background component has a relatively large uncertainty, which will lead to the scatter of $\alpha_{\gamma}$. In the case of $\log {\rm PD}$ versus $\log F_V$, the $p$-value is 0.058 for the linear model, which means that the linear model is marginally true. Assuming the existence of the uncertainty of $P_{\rm max}$, the TCH model is still a hopeful theory to  explain the variation of PD. The possibility that the TCS model with fine-tuned parameters can explain the PD variation is still not ruled out.

\begin{table*}
\begin{center}
\caption{{Results of KS-test and reduced chi-squares  for model fittings}\label{tab:ks}}
\label{test}
\begin{tabular}{lcccccc}
\hline
\hline
 &\multicolumn{2}{c}{$\alpha_{\rm opt}$ vs $\log F_V $} & \multicolumn{2}{c}{$\alpha_{\gamma}$ vs $\log F_{\gamma}$}& \multicolumn{2}{c}{$\log$PD  vs $\log F_V$}\\
\cline{2-3} \cline{4-5} \cline{6-7}
 & TC & linear & TC & linear & TC(Helical)  & linear\\
\hline
Distance        & 0.050 & 0.053 & 0.111 & 0.136 & 0.045  &0.069 \\
$p$-value		& 0.314 & 0.512 & 0.162 & 0.047 & 0.441  &0.058 \\
$\chi^2_{\nu}$  & 1.158 & 1.378 & 4.156 & 4.415 & 1.073  &0.993 \\
\hline
\end{tabular}
\end{center}
{{Here TC(Helical)  denote the model that considering $\Pi_{\rm jet,H}$.} }
\end{table*}

\subsection{Discussion}

The non-linear correlation between $\log$PD and {$\log F_{V}$} for this target was historically studied by \cite{Raiteri:2012}. They claimed that the nonlinear correlation is due to the dilution effect, which is equivalent to our TC model. They estimated the brightness of the non-polarized component as $R_{QSO}=17.85$ magnitude, which roughly agrees with our setup for $F_{V}$. However, they considered that the polarization of the jet is due to the changed viewing angle in the shock in jet model.
The shock in jet model can not be excluded with current constraints, since the variation of jet component, whether it is due to the shock or the Doppler boosting, lead to the same variation behaviors of the PD and color index. The detailed study of profiles of PD and flux light curves can help to further distinguish these two models. Since there is  a significant correlation between optical and radio light curves, it can be inferred that the geometrical setup of the optical
radiation is similar to that of the radio radiation, which has helical trajectories in jet. Thus, the helical jet model is favored to account for the change of the viewing angle. \cite{Zheng:2017} found that the SED of this target is well fitted by the SSC and the EC process with the seed photons from BLR and dust torus. Through  EC process, the seed photons from BLR (keV) can be up scattered by electrons (with Lorentz factor $\gamma\sim 10^3$) to produce GeV photons, which may contribute to the constant component of $\gamma$-ray fluxes.
Similar to the mechanism to explain the optical SWB trend, the SWB trend of the $\gamma$-ray can be understood that the increase of the jet component with a soft spectrum will reduce the spectral index if the constant background has a hard spectrum. The origin of this background component can be studied by the spectral energy distribution from radio to the very high energy bands \citep{Magic:2020}.


\section{Conclusion} \label{Sec:conclusion}

In this work, we collected the decade light curves of
the $\gamma$-ray, optical, radio 15 GHz, and optical PD for FSRQ target B2 1633+382, and investigated the optical and $\gamma$-ray emission regions and the variation mechanism. The principal conclusions are summarized as follows

\begin{itemize}
  \item
Based on the LCCF analysis, the light curves of the optical $R$ band and $\gamma$-ray  are correlated with that of the radio 15 GHz with 3$\sigma$ significance. The $\gamma$-ray  and optical emitting regions are about $14.2_{-2.4}^{+0}$ pc and $14.2_{-8.3}^{+8.3}$ pc upstream of the core region of radio 15 GHz, respectively.  Both of these two regions locate far beyond the BLR.

  \item
Both the optical and $\gamma$-ray spectral indices show the SWB trend. The PD increases with the optical flux. {With certain physical assumptions, the TC model, to some extent, performs better than the linear fitting model. The variation of jet component is probably due to the change of the viewing angle. The shock in jet model is not ruled out by the constraint of this work. }

\end{itemize}

\section*{Acknowledgements}
{We thank the anonymous referee for constructive comments.}
This work has been funded by the National Natural Science Foundation of China under Grant No. U1531105, 11403015, {12063005, and U2031102, and by the Shandong Provincial Natural Science Foundation under grant No. ZR2020MA062}.
Data from the Steward Observatory spectropolarimetric monitoring project were used. This program is supported by Fermi Guest Investigator grants NNX08AW56G, NNX09AU10G, NNX12AO93G, and NNX15AU81G.
The author gratefully acknowledges the optical observations provided by the KAIT.
This research has made use of data from the OVRO 40-m monitoring program (Richards, J. L. et al. 2011, ApJS, 194, 29) which is supported in part by NASA grants NNX08AW31G, NNX11A043G, and NNX14AQ89G and NSF grants AST-0808050 and AST-1109911.

\section*{Data Availability}

The data analysed in this study can be freely retrieved from public archival database. The $\gamma$-ray data are downloaded from the official websites at \url{https://fermi.gsfc.nasa.gov/ssc}(FSSC). The optical spectra, color index, and polarization data are collected from \url{http://james.as.arizona.edu/~psmith/Fermi}(SO). Part of optical $R$-band  data are collected from \url{http://herculesii.astro.berkeley.edu/kait/agn}(KAIT). The radio 15 GHz data are retrieved from \url{http://www.astro.caltech.edu/ovroblazars}(OVRO). The MCMC procedure was taken within the python module {\it emcee} from the website \url{https://emcee.readthedocs.io}\citep{Foreman:2013}.

\end{document}